# $K$-user Interference Channels: Achievable Secrecy Rate and Degrees of Freedom

Xiang He   Aylin Yener
Wireless Communications and Networking Laboratory
Electrical Engineering Department
The Pennsylvania State University, University Park, PA 16802
xxh119@psu.edu   yener@ee.psu.edu

*Abstract*—In this work, we consider achievable secrecy rates for symmetric $K$-user ($K \geq 3$) interference channels with confidential messages. We find that nested lattice codes and layered coding are useful in providing secrecy for these channels. Achievable secrecy rates are derived for very strong interference. In addition, we derive the secure degrees of freedom for a range of channel parameters. As a by-product of our approach, we also demonstrate that nested lattice codes are useful for K-user symmetric interference channels without secrecy constraints in that they yield higher degrees of freedom than previous results.

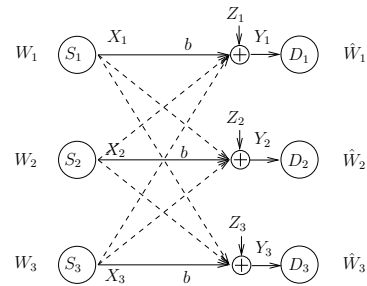

Fig. 1. $K$-User Symmetric Interference Channel, $K = 3$

## I. INTRODUCTION

In a wireless environment, interference is always present. Traditionally, interference is viewed as a harmful physical phenomenon that should be avoided. Yet, from the secrecy perspective, if interference is more harmful to an eavesdropper, it can be a resource to protect confidential messages. To fully appreciate and evaluate the potential benefit of interference to secrecy, the fundamental model to study is the interference channel with confidential messages. This model with two users has been investigated extensively up to date, e.g., [1]–[3].

The $K$-user ($K \geq 3$) interference channel, when all link coefficients are i.i.d. fading, has been studied both with and without secrecy constraints [4], [5]. In these references, the key ingredient for achievability is interference alignment in temporal domain. For the case without secrecy constraints, reference [4] proves the degree of freedom characterization to be $K/2$ for the sum rate.

For the static channel without secrecy constraints, [6] shows the degrees of freedom can not exceed $K/2$, though whether this bound is achievable remains elusive except for when the channel gains of the intended links are algebraic irrational and the other channel gains are rational numbers [7]. References [8], [9] show $K/2$ can be approached asymptotically for a static $K$-user symmetric channel if the channel gain of the interfering link goes to $0$ or $\infty$. Both [8] and [9] employ the idea of interference alignment in the signal space: Reference [8] uses the $Q$-bit expansion and reference [9] uses the lattice code with a sphere as the shaping set [10].

For the static channel with confidential messages, the problem of finding the secure degrees of freedom has largely remained unaddressed so far. In this paper, we focus on the $K$-user ($K \geq 3$) interference channel with confidential messages, where each receiver is an eavesdropper with respect to messages not intended for it. We first derive achievable rates using nested lattice codes for very strong interference. We then investigate the secure degrees of freedom of the sum rate for this channel. We show that positive secure degrees of freedom are achievable, made possible by the fact that users can protect each other via cooperative jamming [11]. Inspired by [9], a layered encoding and decoding scheme is used. The achieved secure degrees of freedom is roughly half of the achievable degrees of freedom in the model without secrecy constraints and is achievable for both weak and strong interference regime. The key ingredient is a tool first introduced in [12] which allows us to bound the secrecy rates under nested lattice codes.

As a by-product of our approach, we also show that for the case without secrecy constraints, a degree of freedom higher than found in [8], [9] is achievable. The main reason leading to this improvement is the use of the nested lattice codes instead of sphere-shaped lattice codes as in [9]. This leads to different decodability conditions and power allocation among different layers.

The rest of the paper is organized as follows: In Section II, we describe the system model. In Section III, we derive the very strong interference condition and the corresponding achievable secrecy rates. Section IV presents the achievable degrees of freedom for the sum rate and the sum secrecy rate and compares it with previous results. Section V concludes the paper.

## II. SYSTEM MODEL

We consider the Gaussian interference channel shown in Figure 1 for $K = 3$. The average power constraint for each source node $S_i$ is $P$. $Z_i, i = 1, ..., K$ are independent Gaussian random variables with zero mean and unit variance. The channel gain coefficient between $S_i$ and $D_i$ is $b$, while the channel gain coefficient between $S_i$ and $D_j$, $i \neq j$ is 1.

Node $S_i$ tries to send a secret message $W_i$ to node $D_i$, while keeping it secret from all the other receiving nodes $D_j, j \neq i$. Hence, for $W_2, ..., W_K$, node $D_1$ is viewed as a potential eavesdropper. Let the signal received by $D_1$ over $N$ channel uses be $Y_1^N$. The corresponding secrecy constraint is given by:

$$\lim_{N \to \infty} \frac{1}{N} H\left(W_2, ..., W_K | Y_1^N\right) = \lim_{N \to \infty} \frac{1}{N} H\left(W_2, ..., W_K\right) \tag{1}$$

The secrecy constraints due to node $D_2, ..., D_K$ are defined in a similar fashion.

## III. ACHIEVABLE SECRECY RATES UNDER VERY STRONG INTERFERENCE

In this section, we summarize several key steps of the achievability proof and derive the very strong interference condition. For clarity, we focus on $K = 3$. The scheme is applicable to $K > 3$ as well.

We note that the achievable scheme is similar to that of the many-to-one interference channel [13]. However, because of the increased connections in the network, the very strong interference condition shall differ from that of [13].

### A. Source Node

Let $(\Lambda, \Lambda_c)$ be a nested lattice structure in $\mathbf{R}^N$, where $\Lambda_c$ is the coarse lattice. The modulus operation $x \mod \Lambda_c$ is defined as $x \mod \Lambda_c = x - arg \min_{y \in \Lambda_c} d(x, y)$, where $d(x, y)$ is the Euclidean distance between $x$ and $y$. The fundamental region $\mathcal{V}(\Lambda_c)$ of the lattice $\Lambda_c$ is defined as the set $\{x : x \mod \Lambda_c = x\}$.

The $i$th source node constructs its input to the channel over $N$ channel uses, $X_i^N$, as follows: Let $t_i \in \Lambda \cap \mathcal{V}(\Lambda_c)$. Let $d_i$ be the dithering noise that is uniformly distributed over $\mathcal{V}(\Lambda_c)$. Then $X_i^N = (t_i^N + d_i^N) \mod \Lambda_c$.

We assume the dithering noise $d_i$ is known by all destination nodes.

### B. Destination Node

Because of the symmetry of the channel, without loss of generality, we focus on the first destination node $D_1$. The destination first decodes the modulus sum of the interference, and then decodes its intended message.

The signal received by $D_1$ over $N$ channel uses is:

$$Y_1^N = bX_1^N + (X_2^N + X_3^N) + Z_1^N \tag{2}$$

Node 1 tries to decode $t_2^N + t_3^N \mod \Lambda_c$. Although $X_1^N$ is not Gaussian, it can be approximated by a Gaussian distribution as $N \to \infty$, as shown in [14, (82)] or [13, (15)-(21)]. Hence we can apply the analysis in [14, Theorem 5], that the probability of decoding error will go to 0 as $N \to \infty$ when

$$R \leq 0.5 \log_2 \left(\frac{1}{2} + \frac{P}{b^2 P + 1}\right) \tag{3}$$

With the knowledge of $t_2^N + t_3^N \mod \Lambda_c$, node 1 can reconstruct $X_2^N + X_3^N \mod \Lambda_c$. After subtracting this term from $Y_1^N \mod \Lambda_c$, the rest part of the interference signal is

$$\left(bX_1^N + Z_1^N\right) \mod \Lambda_c \tag{4}$$

Then, it can be shown [14, (89)] [13, (27)] that if

$$b^2 P + 1 < P \tag{5}$$

then this signal can be approximated by

$$bX_1^N + Z_1^N \tag{6}$$

That is to say:

$$\lim_{N \to \infty} \Pr(bX_1^N + Z_1^N \neq bX_1^N + Z_1^N \mod \Lambda_c) = 0 \tag{7}$$

Finally, the destination tries to decode $t_1$ from (6). Based on [14, Theorem 5], the probability of decoding error will go to zero as $N \to \infty$, if

$$R < C(b^2 P) \tag{8}$$

In summary, if (3), (5) and (8) hold, then the decoding error probability at node 1 should vanish as $N \to \infty$.

### C. Equivocation Rate

The computation of the equivocation rate is the same as [13], as shown below:

$$H\left(t_2^N, t_3^N | Y_1^N, d_i^N, i = 1, 2, 3\right) \tag{9}$$
$$\geq H\left(t_2^N, t_3^N | Y_1^N, X_1^N, Z_1^N, d_i^N, i = 1, 2, 3\right) \tag{10}$$
$$= H\left(t_2^N, t_3^N | X_2^N + X_3^N, d_i^N, i = 1, 2, 3\right) \tag{11}$$

In [12, Theorem 1], it is proved that we can find an integer $T_1, 1 \leq T \leq 2^N$, such that $X_2^N + X_3^N$ is uniquely determined by $\{X_2^N + X_3^N \mod \Lambda_c, T_1\}$. Using this result, (11) equals

$$H\left(t_2^N, t_3^N | X_2^N + X_3^N \mod \Lambda_c, T_1, d_i^N, i = 1, 2, 3\right) \tag{12}$$
$$= H\left(t_2^N, t_3^N | t_2^N + t_3^N \mod \Lambda_c, T_1, d_i^N, i = 1, 2, 3\right) \tag{13}$$
$$= H\left(t_2^N, t_3^N | t_2^N + t_3^N \mod \Lambda_c, T_1\right) \tag{14}$$
$$= H\left(t_2^N, t_3^N | t_2^N + t_3^N \mod \Lambda_c\right) + H\left(T_1 | t_2^N, t_3^N\right)$$
$$\quad - H\left(T_1 | t_2^N + t_3^N \mod \Lambda_c\right) \tag{15}$$
$$\geq H\left(t_2^N, t_3^N | t_2^N + t_3^N \mod \Lambda_c\right) - H(T_1) \tag{16}$$

The first term in (16) can be bounded as follows:

$$H\left(t_2^N, t_3^N | t_2^N + t_3^N \mod \Lambda_c\right) \tag{17}$$
$$= H\left(t_2^N | t_2^N + t_3^N \mod \Lambda_c\right)$$
$$\quad + H\left(t_3^N | t_2^N, t_2^N + t_3^N \mod \Lambda_c\right) = H\left(t_2^N\right) = NR \tag{18}$$

where $R$ is the rate of the codebook computed as $R = \frac{1}{N} \log_2 \|\Lambda \cap \mathcal{V}(\Lambda_c)\|$.

Hence the mutual information leaked to the eavesdropper is bounded as:

$$I\left(t_2^N, t_3^N; Y_1^N, d_i^N, i = 1, 2, 3\right) \leq N(R + 1) \tag{19}$$

Intuitively, this means each pair of users have to pay $R + 1$ in rate to confuse the eavesdropper. Under a symmetric setting,

each user loses $0.5R+0.5$ in rate. This leaves room of $0.5R-0.5$ for each user to send the secret message, which leads to the following theorem:

*Theorem 1:* For any $R, P, b$ such that (3), (5) and (8) hold, a secrecy rate of $[0.5R - 0.5]^+$ is achievable for each user. If

$$b^2 \leq \min\{\frac{P-1}{P}, \frac{\sqrt{P+\frac{1}{16}} - \frac{3}{4}}{P}\} \quad (20)$$

then $R = C(b^2 P)$.

*Remark 1:* Under this condition on $b^2$, it can be verified (3) and (5) become redundant. Hence the secrecy rate is given when $R$ is selected to be $C(b^2 P)$.

*Remark 2:* Reference [15] considers the 3 user symmetric interference channel without secrecy constraints. A different lattice structure is used [10], where $\mathcal{V}(\Lambda_c)$ is replaced by a sphere or a sphere shell. After power normalization, the very strong interference condition of [15] can be expressed as

$$b^2 \leq \frac{\sqrt{P}-1}{P} \quad (21)$$

Comparing (21) with (20), we notice (20) is slightly looser. Hence, using a nest lattice structure allows a slightly wider range of channel parameter under which the channel has very strong interference.

### D. $K > 3$

Theorem 1 can be extended to the case with more than 3 users. In this case, The achievable rate becomes

$$\left[R - \frac{R}{K-1} - \frac{\log_2(K-1)}{K-1}\right]^+ \quad (22)$$

Equation (3) becomes

$$R \leq 0.5 \log_2\left(\frac{1}{K-1} + \frac{P}{b^2 P + 1}\right) \quad (23)$$

Hence, the very strong interference condition (20) becomes

$$b^2 \leq \min\{\frac{P-1}{P}, \frac{\sqrt{P - c + \frac{(c+1)^2}{4}} - \frac{c+1}{2}}{P}\} \quad (24)$$

where $c = \frac{K-2}{K-1}$.

*Remark 3:* It is then interesting to look at the behavior of the secrecy rate when the number of users $K \to \infty$ in the very strong interference channel. From (22), the secrecy rate will converge to $R$. This means the cost of secrecy per user vanishes. A similar phenomenon is also observed in [13] for the many-to-one interference channel.

## IV. SECURE DEGREES OF FREEDOM

In this section, we derive the achievable secure degrees of freedom for a given channel gain $b$. Like [9], a layered lattice structure is used. However, instead of the spherical code, the nested lattice code is used in order to leverage the representation theorem [12] to bound the secrecy rate.

### A. Source Node

Due to symmetry, we focus on source node 1. The transmitted signal is the sum of signals from different layers. The signal from the $i$th layer over $N$ channel uses, $X_1^N$, is given by

$$X_1^N = \sum_{i=1}^M X_{1,i}^N \quad (25)$$

where, like [9], the total number of levels $M$ is to be determined by total power. $X_{1,i}$ is the signal for the $i$th level, which is given by:

$$X_{1,i}^N = (t_{1,i}^N + d_{1,i}^N) \mod \Lambda_{c,i} \quad (26)$$

where $d_{1,i}^N$ is the dithering noise uniformly distributed over $\mathcal{V}(\Lambda_{c,i})$. We assume the dithering noise for each level at each source node is independent from each other. $t_{1,i}^N$ is taken from the Voronoi code book $\Lambda_i \cap \mathcal{V}(\Lambda_{c,i})$, where the variance of $\mathcal{V}(\Lambda_{c,i})$ is chosen to be $P_i$. Let the rate of this codebook be $R_{k,i}$ for the $k$th user.

### B. Destination Node

*1) Strong Interference Regime:* Like [9], we first examine the case where the destination node decodes the interference first, and then decodes the intended signals. The case where the destination decodes the intended signals first can be analyzed in the similar fashion. Due to symmetry, we focus on destination node $D_1$. For the $i$th layer, the destination node decodes the modulus sum of the interference, subtracts it, then decodes the signal from source node $S_1$. Suppose decoding for all layers $j$, $j > i$, are successful, and the modulus operation at layer $j$ incurs negligible distortion for signals at lower layers. Then the remaining signal after subtracting the decoded signals can be approximated by:

$$Y_{1,i}^N = bX_{1,i}^N + X_{2,i}^N + X_{3,i}^N + \sum_{1 \leq j < i}(X_{2,j}^N + X_{3,j}^N) + \sum_{1 \leq j < i}(bX_{1,j}^N) + Z_1^N \quad (27)$$

Define $A_i$ such that

$$A_i = \sum_{1 \leq j < i}(2P_j) + \sum_{1 \leq j < i}(b^2 P_j) + 1 \quad (28)$$

The decoder then decodes $t_{2,i}^N + t_{3,i}^N \mod \Lambda_{c,i}$. The decoding error will decrease exponentially with the dimension of the lattice $N$ if the lattice is designed properly and

$$R_i \leq 0.5 \log_2\left(\frac{P_i}{\frac{P_{I,i}P_{S,i}}{P_{I,i}+P_{S,i}}}\right) \quad (29)$$

where $P_{I,i}$, $P_{S,i}$ are the power of the interference and the signal respectively:

$$P_{I,i} = b^2 P_i + A_i, \quad P_{S,i} = 2P_i \quad (30)$$

This means that

$$R_i \leq 0.5 \log_2\left(\frac{1}{2} + \frac{P_i}{b^2 P_i + A_i}\right) \quad (31)$$

After decoding $(t_{2,i}^N + t_{3,i}^N) \mod \Lambda_{c,i}$, node $D_1$ subtracts $(t_{2,i}^N + d_{2,i}^N + t_{3,i}^N + d_{3,i}^N) \mod \Lambda_{c,i}$ from $Y_{1,i}^N \mod \Lambda_{c,i}$. The signal after the subtraction is given by:

$$\hat{Y}_{1,i}^N = (bX_{1,i}^N + \sum_{1 \leq j < i} \left(X_{2,j}^N + X_{3,j}^N\right) + \sum_{1 \leq j < i} \left(bX_{1,j}^N\right) + Z_1^N) \mod \Lambda_{c,i} \quad (32)$$

When
$$P_i > P_{I,i} \quad (33)$$

$\hat{Y}_{1,i}^N$ can be approximated by

$$bX_{1,i}^N + \sum_{1 \leq j < i} \left(X_{2,j}^N + X_{3,j}^N\right) + \sum_{1 \leq j < i} \left(bX_{1,j}^N\right) + Z_1^N \quad (34)$$

The decoder then decodes $t_{1,i}^N$ from (34). The decoding error will decrease exponentially with the dimension of the lattice $N$ if the lattice is designed properly as in [14] and

$$R_i \leq 0.5 \log_2 \left(\frac{b^2 P_i}{\frac{P'_{I,i} P'_{S,i}}{P'_{I,i} + P'_{S,i}}}\right) \quad (35)$$

where
$$P'_{I,i} = A_i, \quad P'_{S,i} = b^2 P_i \quad (36)$$

This means that
$$R_i \leq 0.5 \log_2 \left(1 + \frac{b^2 P_i}{A_i}\right) \quad (37)$$

The decoder will then subtract $b(t_{1,i}^N + d_{1,i}^N \mod \Lambda_{c,i})$ from (34) and proceed to decode the lower layers.

We next derive the power allocation among different layers. Like [9], we choose $P_i$ such that the right hand sides of (31) and (37) are equal. This means that

$$\frac{1}{2} + \frac{P_i}{b^2 P_i + A_i} = 1 + \frac{b^2 P_i}{A_i} \quad (38)$$

It is easy to check that (38) leads to:
$$P_i = \frac{2 - 3b^2 + \sqrt{4 - 12b^2 + b^4}}{4b^4} A_i \quad (39)$$

For $P_i$ to be a real number, we require $4 - 12b^2 + b^4 > 0$. This, along with the fact that $A_i > 0$, means

$$b^2 < 6 - 4\sqrt{2} \quad (40)$$

which is about 0.34315. Define
$$\alpha = \frac{2 - 3b^2 + \sqrt{4 - 12b^2 + b^4}}{4b^4}, \quad \beta = b^2 + 2 \quad (41)$$

Then $P_1 = \alpha$ and
$$P_i = \alpha \left(\beta \sum_{1 \leq j < i} P_j + 1\right), i > 1 \quad (42)$$

Therefore
$$P_i = \alpha (\alpha\beta + 1)^{i-1} \quad (43)$$

The power expended by each user is given by
$$P = \sum_{i=1}^{M} P_i = \frac{(\alpha\beta + 1)^M - 1}{\beta} \quad (44)$$

Since $\alpha\beta > 0$, we have $\lim_{M \to \infty} P = \infty$.

Under this power allocation, $R_{1,i}$ is given by
$$R_{1,i} = \frac{0.5}{2} \log_2 \left(\frac{1}{2} + \frac{P_i}{b^2 P_i + A_i}\right) + \frac{0.5}{2} \log_2 \left(1 + \frac{b^2 P_i}{A_i}\right) \quad (45)$$

$$= \frac{1}{2} \left(0.5 \log_2 \left(\frac{A_{i+1}}{A_i}\right) - 0.5\right) \quad (46)$$

Therefore $\sum_{i=1}^{M} R_{1,i} = \frac{1}{2} (0.5 \log_2 (A_{M+1}) - 0.5M)$.

Let $R_k$ denote the rate of the $k$ user. Hence, $R_k = \sum_{i=1}^{M} R_{k,i}, k = 1, 2, 3$. If there is no secrecy constraints, the degree of freedom is given by

$$\lim_{P \to \infty} \frac{\sum_{k=1}^{3} R_k}{\frac{1}{2} \log_2 \left(\sum_{i=1}^{3} P_i\right)} = 1.5 - \lim_{M \to \infty} \frac{1.5M}{\log_2 P} \quad (47)$$

$$= 1.5 - \frac{1.5}{\log_2 (\alpha\beta + 1)} \quad (48)$$

Let $R_{e,k}$ denote the rate of the $k$ user. When there are secrecy constraints, each layer can support a secrecy rate of $[0.5R_{1,i} - 0.5]^+$. The secure degrees of freedom is given by

$$\lim_{P \to \infty} \frac{\sum_{i=1}^{3} R_{e,i}}{\frac{1}{2} \log_2 \left(\sum_{i=1}^{3} P_i\right)} \geq \lim_{M \to \infty} \frac{3 \times \left(\sum_{i=1}^{M} (0.5R_{1,i} - 0.5)\right)}{\frac{1}{2} \log_2 (3P)} \quad (49)$$

$$= \frac{3}{4} - \frac{3.75}{\log_2 (\alpha\beta + 1)} \quad (50)$$

We still need to check if the condition (33) are met. Under the current power allocation, we have
$$A_i = (\alpha\beta + 1)^{i-1} \quad (51)$$

(33) means
$$P_i \geq b^2 P_i + A_i \quad (52)$$

Hence we require
$$(1 - b^2)\alpha \geq 1 \quad (53)$$

which holds when (40) holds. In summary, we have the following theorem:

*Theorem 2:* If $b^2 < 6 - 4\sqrt{2}$, the following degrees of freedom for the sum rate is achievable:

$$1.5 - \frac{1.5}{\log_2 (\alpha\beta + 1)} \quad (54)$$

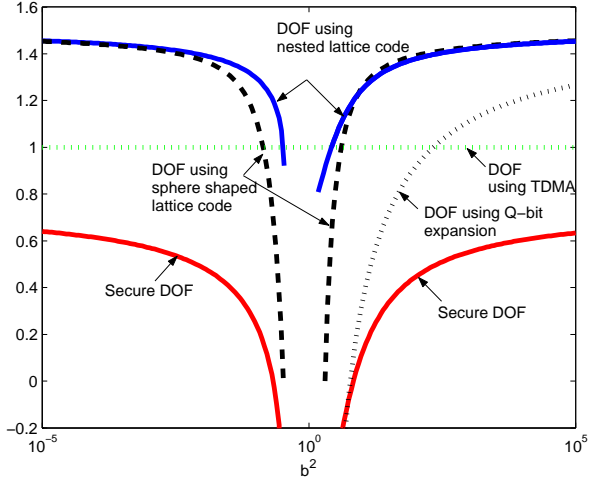

Fig. 2. Degrees of freedom (DOF)

Moreover, the following degrees of freedom for the sum *secrecy* rate is achievable:

$$\left[\frac{3}{4} - \frac{3.75}{\log_2(\alpha\beta+1)}\right]^+ \quad (55)$$

*2) Weak Interference Regime:* When the intended signal is strong enough, the destination should decode it first, and then decode the interference later. In this case, (30) and (36) become

$$P_{I,i} = 2P_i + A_i, \qquad P_{S,i} = b^2 P_i \quad (56)$$
$$P'_{I,i} = A_i, \qquad P'_{S,i} = 2P_i \quad (57)$$

Equation (38) becomes

$$1 + \frac{b^2 P_i}{2P_i + A_i} = \frac{1}{2} + \frac{P_i}{A_i} \quad (58)$$

This means $P_i$ is given by (43) with $\alpha$ given below

$$\alpha = 0.25\left(b^2 + \sqrt{b^4+4}\right) \quad (59)$$

and $\beta$ remains as $b^2 + 2$.

In order for the modulus operation to introduce negligible distortion to lower layers, we require

$$P_i > P'_{I,i} \quad (60)$$

This translates into $\alpha > 1$. This means $b^2 > 3/2$. Hence, we have the following theorem:

*Theorem 3:* If $b^2 > 3/2$, then the degrees of freedom given by (54) and *secure* degrees of freedom given by (55) are achievable, where $\alpha$ is given by (59) and $\beta = b^2 + 2$.

*3) Numerical Results:* As shown in Figure 2, when $b^2 \to \infty$ or $b^2 \to 0$, the secure degrees of freedom converge to $\frac{3}{4}$, which is half the secure degrees of freedom achievable in the model without secrecy constraints.

Also compared in Figure 2 are the degrees of freedom when there are no secrecy constraints. The black dashed lines show the degrees of freedom from [9] using a sphere shaped lattice code. The blue dotted line denotes the degrees of freedom achieved by the $Q$-bit expansion method in [8], which is $\frac{K}{2}(1-\log_b(2K))$, where $K=3$ in Figure 2. The blue lines are the degrees of freedom achieved by our proposed scheme using the nested lattice code. We see that it consistently outperforms the scheme from [9] when $b^2 < 6 - 4\sqrt{2}$ or when $3/2 < b^2 < 8$.

## V. CONCLUSION

In this work, we have considered the symmetric $K$-user ($K \geq 3$) interference channel with or without confidential messages. We have derived the very strong interference condition and the achievable secrecy rates. We have also derived the achievable degrees of freedom for the sum rate and the secrecy sum rate. Both results use nested lattice codes and are shown to outperform previous results. We conclude that nested lattice codes are useful for providing secrecy for $K$-user interference channels with confidential messages, and improve upon the previous constructions in degrees of freedom for $K$-user interference channels without secrecy constraints.